%% file: library.tex
\documentclass[conference]{IEEEtran}
\IEEEoverridecommandlockouts
\usepackage{cite}
\usepackage{amsmath,amssymb,amsfonts}
\usepackage{algorithmic}
\usepackage{graphicx}
\usepackage{tikz}
\usetikzlibrary{tikzmark}
\usetikzlibrary{arrows.meta}
\usepackage[automake,nonumberlist,nogroupskip,nopostdot]{glossaries}

\makeglossaries
\usepackage{textcomp}
\usepackage{multirow}
\usepackage{xcolor}
\usepackage[table]{xcolor}
\usepackage[usenames,dvipsnames]{xcolor}

\usepackage[T1]{fontenc}
\usepackage{listings}
\usepackage{upquote}
\usepackage{enumitem}

\begin{document}
\input{acronym}

% Title for Conference
\title{LLMs' Suitability for Network Security: A Case Study of STRIDE Threat Modeling
}

% % Title for arXiv
% \title{LLMs' Suitability for Network Security: A Case Study of STRIDE Threat Modeling \\ \vspace{0.3cm} 
% \Large \textbf{Author’s draft for soliciting feedback - May 6, 2025}
% % \thanks{Identify applicable funding agency here. If none, delete this.}
% }

\author{\IEEEauthorblockN{AbdulAziz AbdulGhaffar}
\IEEEauthorblockA{\textit{Dept. of Syst. \& Comp. Engineering} \\
%\textit{Computer Engineering} \\
\textit{Carleton University}\\
%Ottawa, Canada \\
abdulazizabdulghaff@cmail.carleton.ca}
\and
\IEEEauthorblockN{Ashraf Matrawy}
\IEEEauthorblockA{\textit{School of Information Technology} \\
\textit{Carleton University}\\
%Ottawa, Canada \\
ashraf.matrawy@carleton.ca}
}

\maketitle
% \tableofcontents
% \newpage
% \clearpage

\begin{abstract}
\gls{ai} is expected to be an integral part of next-generation \gls{ai}-native \acrshort{6g} networks. With the prevalence of \gls{ai}, researchers have identified numerous use cases of \gls{ai} in network security. However, there are very few studies that analyze the suitability of \glspl{llm} in network security. To fill this gap, we examine the suitability of \glspl{llm} in network security, particularly with the case study of \acrshort{stride} threat modeling. We utilize four prompting techniques with five \glspl{llm} to perform \acrshort{stride} classification of \acrshort{5g} threats. From our evaluation results, we point out key findings and detailed insights along with the explanation of the possible underlying factors influencing the behavior of \glspl{llm} in the modeling of certain threats. The numerical results and the insights support the necessity for adjusting and fine-tuning  \glspl{llm} for network security use cases. 
\end{abstract}


\section{Introduction}

% \textcolor{red}{AI-native 6G network, `zero-touch' network, threat identification}

% \textcolor{red}{The main objective is to evaluate the performance of LLM models with various prompting and search techniques and understand the suitability of LLM models for telecommunication tasks. This work is important as it provides insights on using Artificial Intelligence (AI) for next-generation `AI-Native’ and `zero-touch’ telecom networks.}

% \textcolor{red}{Our goal is not to compare LLMs but to study the suitability of LLMs in telecom network use cases. Which applications are these LLMs suitable for? Which tasks will LLM be able to work on?}\\

Future networks, such as \gls{6g} networks, are envisioned to integrate \acrfull{ai} into their networks to be \textit{\gls{ai}-Native} networks~\cite{ericsson_ai_native} to improve performance, efficiency, and scalability~\cite{thomas2024causal}. Ericsson's report~\cite{Ericsson_AI_telecom} indicates that deploying \gls{ai} in telecom networks will not only reduce the \gls{opex} of the network but also provide a $5\%$ to $10\%$ return on investment. On the other hand, with the increasing popularity of \gls{ai} and \acrfullpl{llm}, researchers are identifying potential applications and use cases of \gls{ai} and \glspl{llm} in  networks~\cite{shahid2025large, zhou2024large, ferrag2025generative}. These potential use cases include, but are not limited to, network optimization~\cite{shahid2025large}, automation of security~\cite{zhou2024large}, and threat classification~\cite{ferrag2025generative}.

Upon examining the literature, we notice a significant gap where there is a lack of work analyzing and investigating the suitability of \glspl{llm} in the proposed network security use cases. This motivated us to investigate the suitability of \glspl{llm} in network security use cases. Due to the importance of threat modeling as a starting point in any security exercise, we focus on the ``\textit{\gls{stride}}'' threat model~\cite{Microsoftstride, kohnfelder1999threats}. 

We have extensive experience with \gls{5g} threat modeling using \gls{stride}~\cite{sattar2021stride,mahyoub2024security}. Hence, in this work, we select a case study of \gls{stride} threat modeling to perform \gls{llm}-based classification of \acrshort{5g} threats. We perform the experiments using four prompting techniques with five different \glspl{llm}. This work is important as it provides insights on using \glspl{llm} for threat classification in next-generation `\textit{\gls{ai}-Native}’ telecom networks.

% In light of the aforementioned discussion, we identify the following research questions:
% \begin{itemize}
%     \item \textbf{RQ1:} How suitable are \glspl{llm} for network security use cases?
%     \item \textbf{RQ2:} How well can \glspl{llm} perform \gls{stride} threat modeling in mobile networks?
%     \item \textbf{RQ3:} What are the insights from \gls{llm}-based \gls{stride} classification?
%     % \item \textbf{RQ4:} Which network security tasks will \gls{llm} be able to work on?
%     % \item \textbf{RQ5:} How do \glspl{llm} perform with network threat classification?
%     % \item \textbf{RQ6:} Does using different prompting techniques improve performance?
% \end{itemize}

% \color{black}

% \textcolor{green}{Ashraf: Contributions: Insights and numerical testing. \\
% Most important contribution is the explanation of why LLMs behave in a certain way with certain threats (S T R I D E)}

% To address the research questions, 
The main contributions of this work are as follows:
\begin{enumerate}
    \item We investigate the suitability of \acrfullpl{llm} in network security use cases. For this purpose, we select the case study of \gls{stride} threat modeling of \gls{5g} threats and vulnerabilities. We perform experiments by selecting six \gls{5g} threats and utilizing four different prompting techniques with five \glspl{llm}. 
    \item We provide detailed insights based on the evaluation results of \gls{llm}-based \gls{stride} classification. We present detailed discussions on potential underlying factors that influence the behavior of \glspl{llm} in modeling certain threats, including incorrect threat perspective, failure to identify second-order threats, and insights on \gls{fs} prompting positively impacting performance.
    \item We analyze the suitability of \glspl{llm} using numerical testing and various performance metrics, including accuracy, precision, recall, and F1 score. Our results indicate that the performance of the selected \glspl{llm} is comparable, highlighting the need for enhancements in these models for \gls{stride} threat modeling in \gls{5g} networks.

\end{enumerate}

It should be noted for clarity that our study focuses on classifications and predictions using \glspl{llm} in the context of \gls{stride} threat modeling. Therefore, other performance metrics such as inference speed, adaptability, or scalability of \glspl{llm} are outside the scope of this work.

% our work focuses only on classifications and predictions using \glspl{llm}, rather than examining the inference speed, adaptability, or scalability of these models. 

The paper is organized as follows: Section~\ref{sec:related} provides the motivation of our work in light of the examined related works. Section~\ref{sec:method} explains the evaluation methodology we use for the \gls{stride} threat modeling case study. The detailed results and insights of the evaluation are presented in Section~\ref{sec:result}. Finally, Section~\ref{sec:conclusion} provides the discussion and conclusion of our study.

\section{Motivation and Examination of Related Work} \label{sec:related}
With the advent of \glspl{llm}, many researchers have identified various potential use cases of \glspl{llm} in telecom networks and cybersecurity. We present the most relevant papers in this section.

% \textcolor{green}{Ashraf: Many papers and white papers are suggesting the use of LLM for various security operations, vulnerability checks, etc. Very limited to no word actually testing the suitability of LLMs for these functions. }
% , and performing intent-based management using \glspl{ltm}
\noindent {\bf \gls{llm} for Networking:}  The white paper by Shahid \textit{et al.}~\cite{shahid2025large} presents the concept of ``\gls{ltm}'' for use cases of telecom networks. Some of the potential use cases they mention include the use of \gls{ltm} at the network edge, \gls{ltm} for network optimization, and using \gls{ltm} for network automation tasks, etc. Similarly, Zhou \textit{et al.}~\cite{zhou2024large} identify four different areas of telecom networks that may benefit from \glspl{llm}. These areas include generation, optimization, classification, and prediction problems in telecom networks. Wu \textit{et al.}~\cite{wu2024netllm} present the NetLLM method to utilize \glspl{llm} for three different networking problems, namely, ``viewport prediction'', ``adaptive bitrate streaming'', and ``cluster job scheduling''. The authors extensively evaluate the performance of their proposed framework within these problems.

% verification of security protocols
\noindent {\bf \gls{llm} for Security:} Ferrag \textit{et al.}~\cite{ferrag2025generative} propose multiple applications of \glspl{llm} in cybersecurity, including detection and analysis of threats, incident response, automation of security tasks, etc. The work by Guthula \textit{et al.}~\cite{guthula2023netfound} proposes a foundation model for security that takes into account the distinct nature of network traffic. The aim of the authors is to consider the general applicability of the model.

S{\k{e}}dkowski~\cite{skedkowski2024threat} studied the efficacy of \glspl{llm} in recognizing potential threats in the network and recommending countermeasures. Their methodology includes using Nmap reports to classify threats using \gls{stride} threat modeling with three different \glspl{llm}. The author concludes that the performance of \glspl{llm} in threat detection is comparable to that of humans. The scope of their work is focused on testing the application of \gls{ai} in threat modeling, as opposed to our aim of studying the suitability of \glspl{llm} for network security.

 % On the other hand, 
Yang \textit{et al.}~\cite{yang2024threatmodeling} present ``ThreatModeling-LLM'', an \gls{llm}-based method to perform threat modeling of the banking system using the \gls{stride} model. Their approach includes various steps to improve the performance of the \glspl{llm}. Their scope is limited to banking systems. The main objective of their work is to improve and automate threat modeling using \glspl{llm}, instead of investigating the suitability of \glspl{llm} for network security. Saha \textit{et al.}~\cite{saha_llm} developed ``ThreatLens'' to perform threat modeling and test plan generation for hardware security verification.

\noindent {\bf Motivation:} After examining the related works, we identify that the suitability of the potential use cases of \glspl{llm} needs to be investigated. However, to the best of our knowledge, there are very few existing studies that are actually testing the suitability of \glspl{llm} for these use cases. This motivated us to study the suitability of the \glspl{llm} in telecom network security using a case study of \gls{stride} threat modeling.

\section{Evaluation Methodology for A Case Study of \acrshort{stride} Threat Modeling} \label{sec:method}

The aim of this case study is to employ \glspl{llm} to categorize the threats and vulnerabilities on \gls{5g} interfaces using the \gls{stride} model. The main objective is to evaluate various prompting and search techniques and investigate the suitability of \glspl{llm} for telecommunication tasks. 

% The main objective is to evaluate various prompting and search techniques and understand the suitability of \glspl{llm} for telecommunication tasks. 

\begin{figure}
    \centering
    \includegraphics[width=0.65\linewidth]{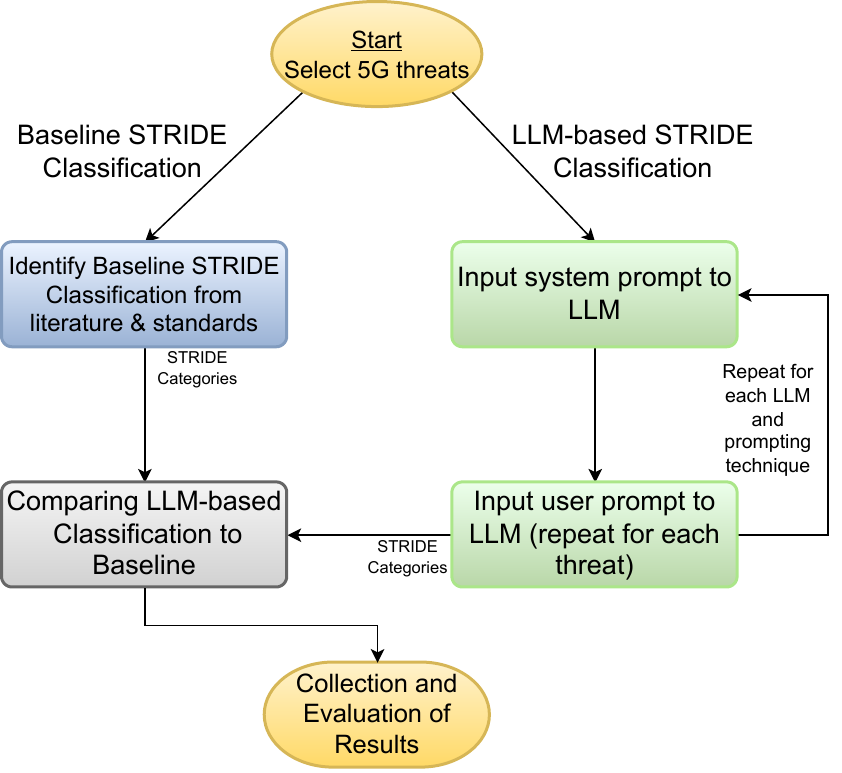}
    \caption{Evaluation Methodology of \acrshort{stride} Modeling}
    \label{fig:methodology}
\end{figure}

Our evaluation methodology is shown in Figure~\ref{fig:methodology}. We initially select multiple threats and vulnerabilities in the \gls{5g} network, along with their baseline \gls{stride} classifications from the published literature and standards. Then, we use various prompting techniques to perform the \gls{llm}-based \gls{stride} classification of the selected threats. Finally, we compare the \gls{stride} classification by \glspl{llm} to the baseline and evaluate the results. We will explain each step in the following.

% with different \glspl{llm} to categorize the threats using the \gls{stride} threat model.

\subsection{Selected \gls{5g} Threats and Vulnerabilities} \label{subsec:threats}
To carry out the evaluation, we select six threats and vulnerabilities on \gls{5g} interfaces from our previous research work~\cite{mahyoub2024security} along with their \gls{stride} classifications as a baseline; this step is shown with a blue block in Figure~\ref{fig:methodology}. As we already mentioned in our previous work~\cite{mahyoub2024security}, we want to clarify that the baseline \gls{stride} classification of the threats may not be unique. We select these attacks from our previously published study~\cite{mahyoub2024security} while ensuring that the selected threats cover the six \gls{stride} categories and span over multiple \gls{5g} interfaces. Three of the selected threats are on the N1 interface, because N1 is exposed to the \gls{ran} and it faces the largest number of threats~\cite{mahyoub2024security}.
In our earlier work~\cite{mahyoub2024security}, we comprehensively explored and identified the threats and vulnerabilities on the critical \gls{5g} network interfaces and categorized them based on the \gls{stride} threat model. The selected threats, along with their description, are explained below. We use the same threat names in the first row of Table~\ref{tab:stride_table}:

% , and the threats may have a different classification depending on the network or threat context. We mainly focus on the classification that is more suitable in the context of \gls{5g} networks. 

\begin{itemize}
    \item \textit{\gls{amf} Impersonation on N1 interface:} If a malicious actor is impersonating \gls{amf}, it can access sensitive user information through the N1 interface~\cite{3GPP.33.501,3GPP.33.926}. This is especially important when users send their unique identifiers (e.g., \gls{supi}) to the \gls{amf} to join the \gls{5g} network~\cite{mahyoub2024security}.
    
    \item \textit{\gls{5g}-\gls{guti} and \gls{imei} correlation on N1 interface:} If the attacker is able to correlate \gls{5g}-\gls{guti} and \gls{imei} of a user, it can trace the present and future mobility and position of the user~\cite{mahyoub2024security, nist_5g}.
    
    \item \textit{Bidding down on Xn-handover:} In this threat, insecure algorithms are enforced by the malicious \gls{gnb} in the \gls{5g} system, resulting in weakening the security of the \gls{5g} system~\cite{mahyoub2024security, 3GPP.33.926}.
    % \textcolor{red}{Clear in D.2.2.6 in 3GPP TR 33.926 version 16.5.0 Release 16}
    
    \item \textit{Eavesdropping on F1 interface:} On the F1 interface, eavesdropping of the control plane and data plane traffic is a potential threat~\cite{mahyoub2024security}. This eavesdropping will result in information disclosure and can further lead to threats that can allow spoofing and tampering as well~\cite{3GPP.33.926, 3GPP.33.824}.
    
    % \textcolor{red}{Threat description from: The spoofing with Eavesdropping is mentioned Clear in 5.3.3.7 3GPP TR 33.926 version 16.5.0 Release 16 //  Clear in 3GPP TR 33.824 V17.0.0 (2022-03)}
    
    \item \textit{False \gls{s-nssai} on N1 interface:} Providing incorrect \gls{s-nssai} during the \gls{nssaa} procedure threatens system resources and may result in escalation of privileges~\cite{3GPP.33.501, 3GPP.33.926}.

    % \textcolor{red}{Threat description from: NSSAA in ENISA THREAT LANDSCAPE FOR 5G NETWORKS, 3GPP TS 33.501 version 17.5.0 Release 17, Clear in 3GPP.33.926}
    
    \item \textit{\gls{mitm} attack on N3 interface:} The N3 interface between \gls{5g} \gls{ran} and \gls{upf} is susceptible to \gls{mitm} attack~\cite{mahyoub2024security}.
\end{itemize}

\subsection{Large-Language Models (LLMs)}
We select the following \acrfullpl{llm} to perform this evaluation: Sonar by Perplexity~\cite{sonar}, GPT-4o by OpenAI~\cite{gpt4o}, Claude 3.7 Sonnet by Anthropic~\cite{sonnet}, Grok-2 by xAI~\cite{grok-2}, and Gemini 2.5 Pro by Google~\cite{gemini}. We use these \glspl{llm} through the pro version of the Perplexity AI platform that we have access to through our University~\cite{perplexity}. We are interested in evaluating the suitability of the current \glspl{llm} for network security, hence we use the \glspl{llm} with base knowledge as it is, without retraining or fine-tuning on any datasets. We perform the \gls{stride} classification of the six selected threats using these \glspl{llm} in order to evaluate the suitability of the \glspl{llm} for network security. Figure~\ref{fig:methodology} shows the \gls{llm}-based \gls{stride} classification methodology with green blocks.

% We did not train or fine-tune any \gls{llm} on any dataset. Instead, we use the \glspl{llm} with base knowledge as it is, without any further training or fine-tuning.

\subsection{\gls{llm} Prompts}
We use a combination of system and user prompts to perform the \gls{llm}-based \gls{stride} classification. 

%\subsubsection{\textbf{System Prompt}}
\noindent{\textbf{System Prompt:}}
The experiment is performed by providing a system prompt to the \gls{llm} at the beginning, which is an instruction to define the scope of the \gls{llm} task and control the output of the \gls{llm}~\cite{hui2024pleak}. The system prompt we provide to the \glspl{llm} is shown in Listing~\ref{lst:sys}. We initially define the \textbf{scope} of the \gls{llm} task and then outline the instructions to refine the \textbf{output} of the \gls{llm}.

% xleftmargin=4pt, breaklines=true, float - The text in \textbf{\textcolor{blue}{Blue}} color defines the scope \\of the \gls{llm} task. While the text in \textbf{\textcolor{cyan}{Cyan}} color refines the output of the \gls{llm}

% \begin{lstlisting}[float,  linewidth=0.90\columnwidth, breaklines=true, caption={\gls{llm} System Prompt. The text in \textbf{\textcolor{blue}{Blue}} color defines the scope of the \gls{llm} task. While the text in \textbf{\textcolor{cyan}{Cyan}} color refines the output of the \gls{llm}}, label={lst:sys}]
\begin{lstlisting}[float=b, linewidth=1\columnwidth, xleftmargin=0.01\columnwidth, xrightmargin=0.01\columnwidth, basicstyle=\scriptsize\ttfamily, caption={\gls{llm} System Prompt. \textbf{\textcolor{blue}{Blue}} text defines the scope of the \\ \gls{llm} task, whereas \textbf{\textcolor{cyan}{Cyan}} text refines the output of the \gls{llm}}, label={lst:sys}]
^You are a 5G network security expert.^
^Your task is to classify a given 5G threat or^ ^vulnerability according to the STRIDE model:^
^1. Spoofing^ ^2. Tampering^ ^3. Repudiation^ ^4. Information disclosure^ ^5. Denial of service^ ^6. Elevation of privilege^
**Each threat or vulnerability may belong to one or more** **STRIDE categories. Your response should list only the** **applicable category or categories without any additional** **details or explanations.**
\end{lstlisting}

%\subsubsection{\textbf{User Prompts}}
\noindent{\textbf{User Prompts:}}
The user prompts are a set of prompts that we run for each of the selected six threats. We use the following two prompting approaches in our evaluation: 

% \begin{itemize}
% \item Prompting Approaches:
\begin{enumerate}
    \item  \textbf{\acrfull{zs} prompting:} The \gls{zs} prompting approach only provides the \gls{llm} with a description of the task without including any examples in the prompt~\cite{brown2020language}. The \gls{zs} prompt we use in this case study is shown in Listing~\ref{lst:user1}.
    \item \textbf{\acrfull{fs} prompting:} This approach of prompting the \gls{llm} includes a certain number of examples in the prompt~\cite{brown2020language}. The \gls{fs} prompt we use is shown in Listing~\ref{lst:user2}.

\end{enumerate}
% \end{itemize}

We replace the \textcolor{red}{`[NAME\_OF\_THREAT]'} in these prompts with the name of a specific threat and provide the prompt to the \gls{llm}. Then, we record the \gls{stride} classification of a threat provided by the \gls{llm}. We repeat this step for all six selected threats and use the same prompts for each threat as shown in Listings~\ref{lst:user1} and~\ref{lst:user2}, for each of \gls{zs} and \gls{fs} prompting techniques, respectively (see steps in green blocks in Figure~\ref{fig:methodology}).

We combine \gls{zs} and \gls{fs} prompting approaches with `base' \gls{llm} knowledge (no internet access) and with `internet' search, to come up with four prompting techniques, \textit{\gls{zs}\_Base}, \textit{\gls{zs}\_Internet}, \textit{\gls{fs}\_Base}, and \textit{\textit{\gls{fs}\_Internet}}. Apart from Zero-shot (ZS) and Few-shot (FS) promptings, another approach that is generally used to redefine the scope of the \glspl{llm} is `fine-tuning'~\cite{brown2020language}. However, fine-tuning requires retraining the \gls{llm} with a specific dataset in order to refine its output for a particular use case. Since this approach is expensive (in terms of time and resources), we do not consider \gls{llm} fine-tuning for the evaluation in this work.

%The user prompts for `Zero-shot (ZS) prompting’ and `Few-shot (FS) prompting’ are provided in Listings~\ref{lst:user1} and~\ref{lst:user2}, respectively. 

% xleftmargin=4pt,

\begin{lstlisting}[float=b, xleftmargin=4pt, linewidth=1\columnwidth, xleftmargin=0.01\columnwidth, xrightmargin=0.01\columnwidth, basicstyle=\scriptsize\ttfamily, caption={User prompt for Zero-Shot (ZS) prompting}, label={lst:user1}]
Classify the following threat/vulnerability:
@[NAME_OF_THREAT]@
\end{lstlisting}

 % xleftmargin=4pt, 

\begin{lstlisting}[float=b, xleftmargin=4pt,linewidth=1\columnwidth, xleftmargin=0.01\columnwidth, xrightmargin=0.01\columnwidth, basicstyle=\scriptsize\ttfamily, caption={User prompt for Few-Shot (FS) prompting}, label={lst:user2}]
Classify the following threat/vulnerability:
@[NAME_OF_THREAT]@
The following are some examples of threat STRIDE classification. Here, {X} represents that the threat does not belong to this category, and {O} means the threat belongs to this category:
1.	NAS protocol-based attack on N1 interface: S{X}, T{X}, R{X}, I{O}, D{O}, E{X}
2.	A bidding down of Security features on N1 interface: S{X}, T{O}, R{X}, I{O}, D{O}, E{X}
3.	Keystream reuse on Xn interface:  S{X}, T{X}, R{X}, I{O}, D{X}, E{X}
4.	Flawed Validation of Client Credentials Assertion on SBI interface: S{O}, T{X}, R{X}, I{O}, D{O}, E{O}

\end{lstlisting}

\begin{table*}
    \centering
    \caption{\acrfull{llm}-based \gls{stride} Classification of \gls{5g} Threats}
    \includegraphics[width=0.8\linewidth]{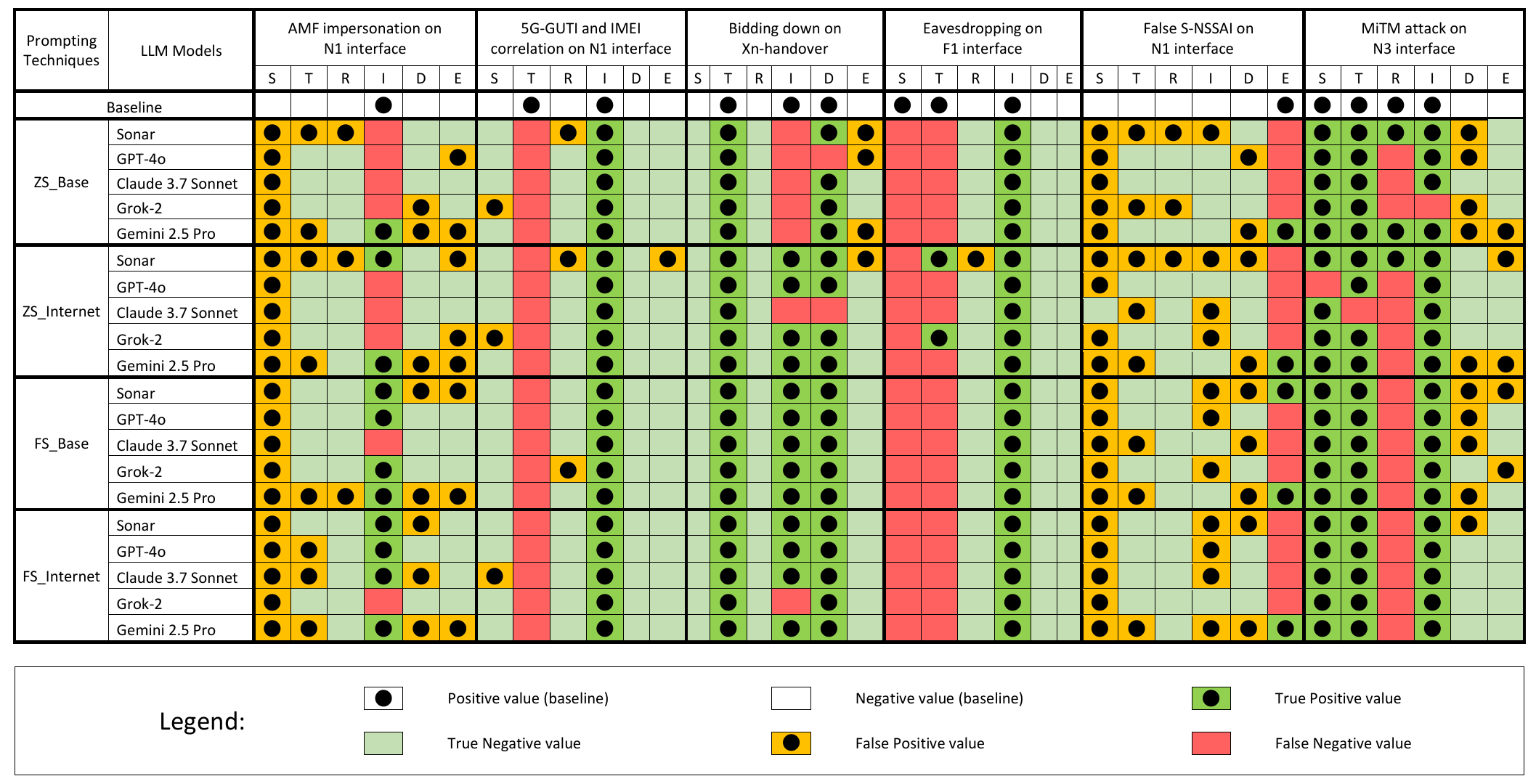}
    \label{tab:stride_table}
\end{table*}

\section{Results and Insights} \label{sec:result}
While we report some performance metrics on how the \glspl{llm} performed in our experiments, we note that our main goal is not to compare the \glspl{llm} but rather to study the suitability of these models for network security tasks using a case study of 5G \gls{stride} modeling. Most importantly, we provide insights on their use in the modeling process. 
% The results of the experiments are shown in the figures.
 %As we explain in Section~\ref{sec:llm},  the \glspl{llm} have a comparable performance when it comes to certain metrics.
% We only present these results here to show the performance of \glspl{llm} relative to each other.

%\vspace{-0.09cm}
\subsection{\gls{llm}-based \gls{stride} Classification}
In this section, we provide the results of evaluating the case study of \gls{stride} threat modeling of \gls{5g} threats. The results of \gls{llm}-based \gls{stride} classification of the six \gls{5g} threats are shown in Table~\ref{tab:stride_table}. The first column of the table includes the prompting techniques we employ in our evaluation, while the second column shows the \glspl{llm} we select. The rest of the columns in this table present the baseline \gls{stride} classification along with the results of \gls{llm}-based \gls{stride} classification of the \gls{5g} threats. In this table, the white cells with a dot $(\bullet)$ represent a positive baseline value, while an empty white cell represents a negative baseline value. According to this baseline, we categorize the \gls{llm}-based \gls{stride} classifications as True Positive (TP) (dark green cell with a dot $(\bullet)$), True Negative (TN) (empty green cell), False Positive (FP) (yellow cell with a dot $(\bullet)$), and False Negative (FN) (empty red cell). The coloring scheme represents that the greens (TP and TN) are correct classifications. Yellow (FP) is an incorrect classification, but it is not the worst outcome (over-predicting positive), and red (FN) is an incorrect classification and represents the worst outcome (under-predicting positive). In the following, we present the main insights and observations from these results.

\subsection{Insights on the \gls{llm}-based \gls{stride} Classification}
\noindent \textbf{Incorrect Threat Perspective:} Looking at the results of the first threat, ``\gls{amf} impersonation on N1 interface'', in Table~\ref{tab:stride_table}, we observe that the \glspl{llm} did not consistently categorize this threat as `information disclosure', similar to the baseline classification (see red cells in column I). We note that this threat is categorized as `spoofing' by all \glspl{llm} with all prompting techniques (yellow cells in column S). This could be attributed to the \glspl{llm} classifying this threat from the perspective of the \gls{amf}, while the baseline classification is from the perspective of the user of the \gls{5g} network. Hence, according to the baseline classification, this threat will only lead to `information disclosure' of the sensitive user information to the malicious \gls{amf}, as described in Section~\ref{subsec:threats}. 

The classification results of the threat, ``False \gls{s-nssai} on N1 interface'', demonstrate a degree of consistency across all prompting techniques and \glspl{llm}. This threat is outlined in 3GPP TR 33.926~\cite{3GPP.33.926} and the 3GPP specified `elevation of privilege' as the corresponding threat category. However, no \gls{llm} correctly identified this threat in all prompting techniques, except Google's Gemini 2.5 Pro, which correctly identified this threat in the `elevation of privilege' category with all prompting techniques (green box with a dot in column E). Furthermore, in almost all cases, the \glspl{llm} incorrectly identified this threat in a `spoofing' category (yellow cells in column S), which is incorrect compared to the 3GPP's categorization in~\cite{3GPP.33.926}. Similar to the first threat, it is very likely that \glspl{llm} consider an incorrect threat perspective and categorize this threat as a `spoofing' attack due to the transmission of false \gls{s-nssai}, instead of an `elevation of privilege' threat.
 % specifically in the context of \gls{5g}, 

\noindent \textbf{Failure to Identify Second-order Threats:} One major observation we notice in Table~\ref{tab:stride_table} is that the fourth threat, ``Eavesdropping on F1 interface'', is only categorized as `information disclosure' and not as `spoofing' and `tampering' in almost all \gls{llm}-based \gls{stride} classifications. This could be because \glspl{llm} are not considering the possible `second-order effect' or `second-order threat' of this attack. However, as specified by 3GPP~\cite{3GPP.33.926, 3GPP.33.824}, due to the lack of confidentiality and integrity measures, the eavesdropping threat may not only result in `information disclosure' but may also result in `spoofing' and `tampering' threats as well. This shows that \glspl{llm} may not always provide a comprehensive threat modeling, specifically when multiple subsequent threats are also possible.

We further see similar behavior in the second selected threat, ``\gls{5g}-\gls{guti} and \gls{imei} correlation on N1 interface'', where the threat is identified correctly as `information disclosure'. This `information disclosure' can further lead to `tampering', but it is not categorized as a `tampering' threat by the \glspl{llm}. On the positive side, the classification performance of the second threat is very consistent across all prompting techniques and \glspl{llm}, and it is slightly improved as we move from the \textit{\gls{zs}\_Base} to \textit{\gls{fs}\_Internet} prompting techniques. 

For the ``\gls{mitm} attack on the N3 interface'' (sixth threat), the \glspl{llm} mostly identified this threat correctly in the `spoofing', `tampering', and `information disclosure' categories. However, they predominantly failed to identify the \gls{mitm} threat in the `repudiation' category. This is similar to the previous results of the fourth threat, where the \glspl{llm} did not identify some categories when multiple subsequent threats are also possible. In the case of a \gls{mitm} attack, for example, it is possible that an attacker may intercept and modify the content of a packet in transit from the sender to the receiver, if appropriate security measures are not provided. The sender will deny the transmission of the modified content. Nonetheless, due to the lack of security measures, it will be difficult to identify the packet modification or the entity responsible for the modification.

\noindent \textbf{\gls{fs} Prompting Improves Performance:} We discover that the classification performance of the third threat, ``Bidding down on Xn-handover'', is mostly accurate in all prompting techniques, except \textit{\gls{zs}\_Base}. For example, the \textit{\gls{fs}\_Base} prompting approach achieved an accuracy of $100\%$ with all \glspl{llm}. This can be due to the fact that the user prompt we provide for \gls{fs} prompting includes a similar example of a `bidding down of Security features on N1 interface' with the same \gls{stride} classification as this threat. We also notice that the performance is increased in \gls{fs} prompting compared to \gls{zs} prompting, similar to the first and second threats.

Similarly, we notice from the first threat that the classification performance is improved as we move from \textit{\gls{zs}\_Base} to \textit{\gls{fs}\_Internet} prompting technique (more green and less red as we move down the `I' column). This also suggests that providing examples to the \glspl{llm} improves their performance.

% \vspace{-1cm}

\subsection{Comparison of Prompting Techniques}
\begin{figure}[!t]
    \centering
    \includegraphics[width=0.8\linewidth]{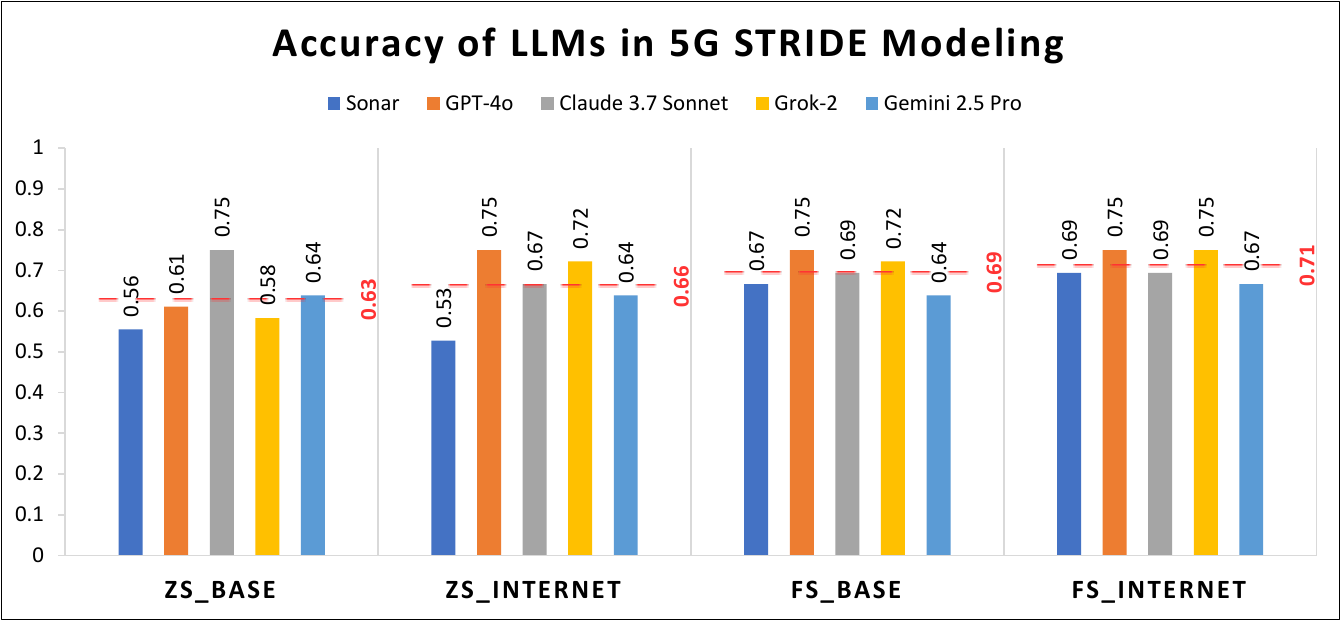}
    \caption{Accuracy of LLMs in 5G STRIDE Modeling using different prompting techniques}
    \label{fig:accuracy}
\end{figure}

\begin{figure}[t]
    \centering
    \includegraphics[width=0.8\linewidth]{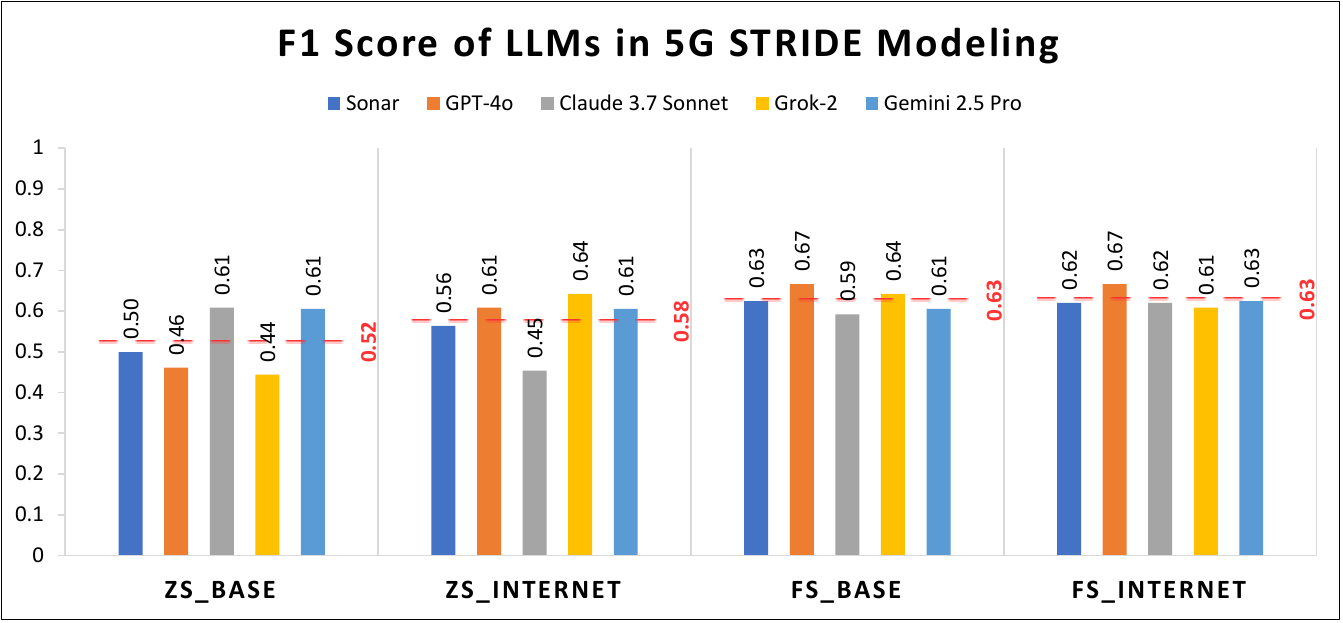}
    \caption{F1 Score of LLMs in 5G STRIDE Modeling using different prompting techniques}
    \label{fig:f1}
\end{figure}

We analyze and compare the performance of the prompting techniques in terms of accuracy and F1 score. The results in Figures~\ref{fig:accuracy} and~\ref{fig:f1} are averaged over the six selected threats ($36$ cells in one row). Figure~\ref{fig:accuracy} shows the accuracy of \glspl{llm} in \gls{5g} \gls{stride} threat modeling with four different prompting techniques. We see from the figure that with \textit{\gls{zs}\_Base} prompting, the accuracy achieved is the lowest compared to other prompting techniques. As we go from \textit{\gls{zs}\_Base} prompting to \textit{\gls{fs}\_Internet} prompting, we observe that the performance (accuracy) is increasing gradually. We plot the average accuracy of each prompting technique and notice that as we move from \textit{\gls{zs}\_Base} to \textit{\gls{fs}\_Internet}, the average accuracy of \glspl{llm} in \gls{5g} \gls{stride} threat modeling is increasing from $63\%$ to $71\%$. These results are in accordance with the evaluation performed by Brown \textit{et al.}~\cite{brown2020language} and their conclusion that providing examples in the prompts improves the performance.

% We analyze and compare the performance of the prompting techniques in terms of accuracy and F1 score. The results in Figures~\ref{fig:accuracy} and~\ref{fig:f1} are averaged over the six selected threats ($36$ cells in one row). Figure~\ref{fig:accuracy} shows the accuracy of \glspl{llm} in \gls{5g} \gls{stride} threat modeling with four different prompting techniques. We see from the figure that with \textit{\gls{zs}\_Base} prompting, Claude 3.7 Sonnet is outperforming other \glspl{llm} with an accuracy of $75\%$. Since no examples are provided in this prompting technique and the base \gls{llm} knowledge is used, this suggests that Claude 3.7 Sonnet has better training on telecom knowledge. As we go from \textit{\gls{zs}\_Base} prompting to \textit{\gls{fs}\_Internet} prompting, we observe that the performance is increasing gradually with other \glspl{llm}, more prominently, GPT-4o and Grok-2. We plot the average accuracy of each prompting technique and notice that as we move from \textit{\gls{zs}\_Base} to \textit{\gls{fs}\_Internet}, the average accuracy of \glspl{llm} in \gls{5g} \gls{stride} threat modeling is increasing from $63\%$ to $71\%$. These results are in accordance with the evaluation performed by Brown \textit{et al.}~\cite{brown2020language} and their conclusion that providing examples in the prompts improves the performance.

% Figure~\ref{fig:precision} illustrates the precision in \gls{5g} \gls{stride} threat modeling. We notice a similar performance trend to the previous results, where  Claude 3.7 Sonnet performs the best with \textit{\gls{zs}\_Base} prompting, while the performance of other \glspl{llm} is improved with \textit{\gls{fs}\_Internet} prompting. Furthermore, the average precision also increases from $54\%$ with \textit{\gls{zs}\_Base} to $65\%$ with \textit{\gls{fs}\_Internet} prompting technique. 

The F1 score of the \glspl{llm} with different prompting techniques is shown in Figure~\ref{fig:f1}. With the \textit{\gls{zs}\_Base} prompting, we observe that the lowest average F1 score is recorded with $52\%$, as compared to the other prompting techniques. We notice that as we go from \textit{\gls{zs}\_Base} to \textit{\gls{fs}\_Internet} prompting technique, the F1 score of the \glspl{llm} increases slightly. This is evident from the average line (shown in red), which shows a $10\%$ increase in the average. This slight increase indicates that the \gls{fs} prompting techniques improve the performance. On the other hand, the performance of the \glspl{llm} relative to each other is almost similar and shows no significant difference.

% The F1 score of the \glspl{llm} with different prompting techniques is shown in Figure~\ref{fig:f1}. With the \textit{\gls{zs}\_Base} technique, the Sonar, GPT-4o, and Grok-2 \glspl{llm} show comparable performance, while Claude 3.7 Sonnet and Gemini 2.5 Pro have similar F1 score. We notice that as we go from \textit{\gls{zs}\_Base} to \textit{\gls{fs}\_Internet} prompting technique, the F1 score of the \glspl{llm} increases slightly. This is evident from the average line (shown in red), which shows a $10\%$ increase in the average. This slight increase indicates that the \gls{fs} prompting techniques improve the performance. On the other hand, the performance of the \glspl{llm} relative to each other is almost similar and shows no significant difference.

% \begin{figure}
%     \centering
%     \includegraphics[width=0.9\linewidth]{Figures/Graph_Precision.pdf}
%     \caption{Precision of LLMs in 5G STRIDE Modeling using different prompting techniques}
%     \label{fig:precision}
% \end{figure}

% \begin{figure}
%     \centering
%     \includegraphics[width=\linewidth]{Figures/Graph_Recall.pdf}
%     \caption{Recall of LLMs in 5G STRIDE Modeling using different prompting techniques}
%     \label{fig:recall}
% \end{figure}

% \begin{figure}
%     \centering
%     \includegraphics[width=\linewidth]{Figures/Graph_Comparison.pdf}
%     \caption{Comparison of different LLMs in 5G STRIDE Modeling}
%     \label{fig:comparison}
% \end{figure}

\subsection{Performance of \glspl{llm} in \gls{stride} Classification} \label{sec:llm}
\begin{figure}
    \centering
    \includegraphics[width=0.8\linewidth]{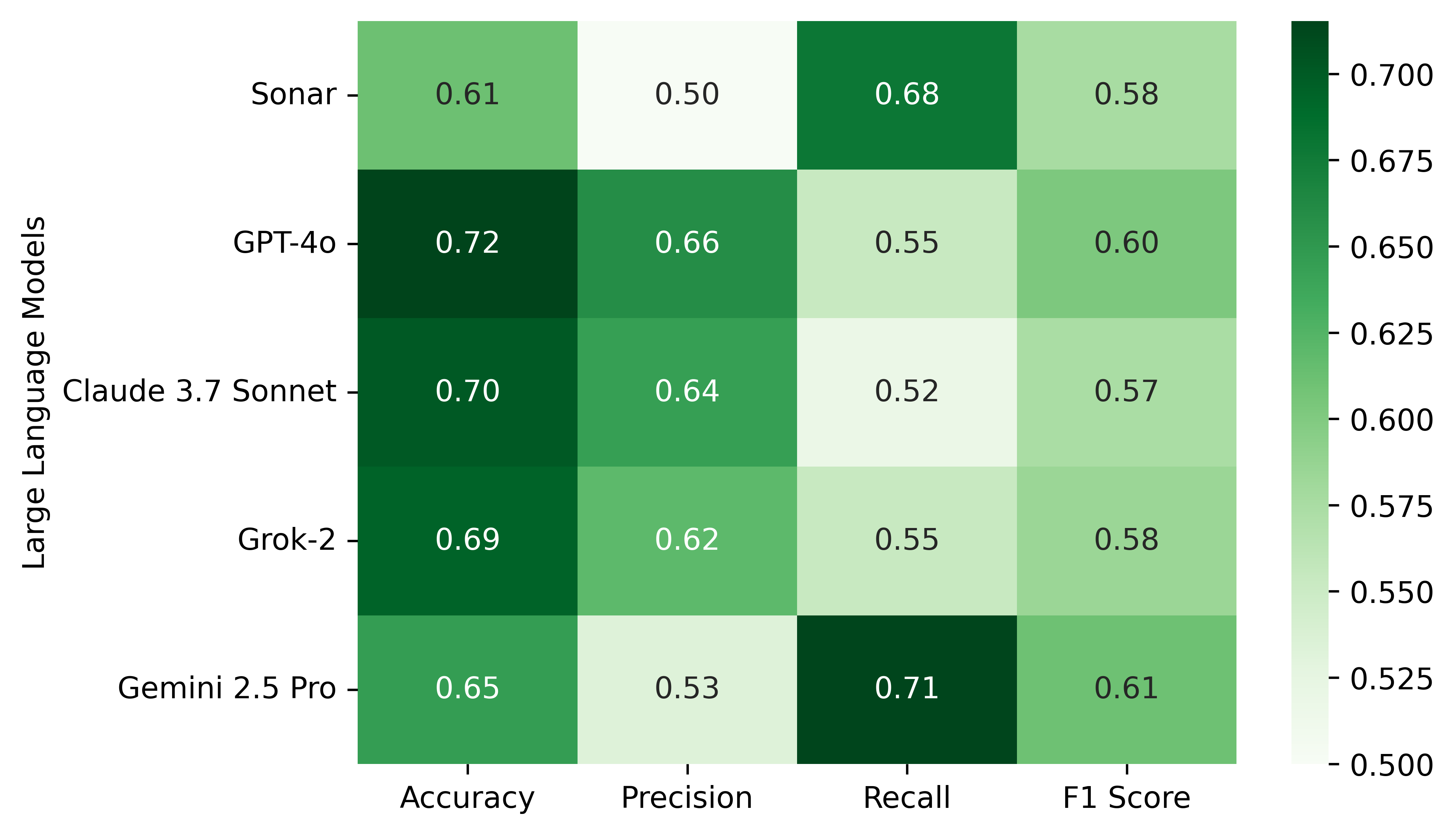}
    \caption{Heatmap showing the performance of \glspl{llm} in terms of accuracy, precision, recall, and F1 score. Higher values (dark green color) are better.}
    \label{fig:heatmap}
\end{figure}

 The results show that \glspl{llm}' performance in our experiments is mostly comparable. Figure~\ref{fig:heatmap} illustrates the performance of the \glspl{llm} in terms of accuracy, precision, recall, and F1 score using a heatmap chart. This result is averaged across all threats and all prompting techniques for a specific \gls{llm}. We observe that GPT-4o, Claude 3.7 Sonnet, and Grok-2 show `relatively' higher accuracy and precision but lower recall compared to the other two \glspl{llm}. On the other hand, Sonar and Gemini 2.5 Pro achieved lower precision and higher recall in comparison. Higher recall means that these models capture most of the TP cases and keep FN (the worst outcome) to a minimum. The F1 score indicates that the performance of all the \glspl{llm} is comparable for this case study. The maximum accuracy achieved is $72\%$, which highlights that there are opportunities for improvement across all \glspl{llm}, perhaps by fine-tuning for the specific application of \gls{stride} threat modeling in \gls{5g} networks.

 % of Our Work
\begin{itemize}[leftmargin=*]
    \item \textbf{Limitations:} As we already mentioned in Section \ref{subsec:threats}, the baseline \gls{stride} classification of the threats may not be unique. However, we are more interested in investigating the behavior of \glspl{llm} and identifying the insights on the \gls{llm}-based \gls{stride} classification, rather than focusing on the accuracy of individual \gls{llm} classification.

    \item \textbf{Challenges:} We note that several articles highlight the prominent challenges with the \glspl{llm} \cite{huang2024large,boateng2025survey}. The most relevant issues to threat modeling include incorrect predictions and \gls{llm} hallucinations, which may result in disregarding required countermeasures or implementing unnecessary security measures. Secondly, the adaptability of \glspl{llm} for telecom-specific threats and vulnerabilities. Thirdly, improving \gls{llm} inference speed in networks to enable rapid threat modeling of the detected threats. 
\end{itemize}

% and issues 

% width=0.9\linewidth

\section{Discussion and Conclusion} \label{sec:conclusion}

In this work, we explore and investigate the suitability of \glspl{llm} for network threat modeling. To perform this analysis, we select the case study of \acrshort{stride} threat modeling to perform \gls{llm}-based classification of \acrshort{5g} threats according to the \gls{stride} threat model. We observe from our evaluation that providing examples to the \glspl{llm} using \gls{fs} prompting improves their performance. We further notice that \glspl{llm} may not always consider the threat classifications as a result of the second-order effect. This will limit the threat identification and may eventually result in not identifying all the possible risks associated with a threat. We hope these insights and results of our work are the starting points to encourage research into fine-tuning \glspl{llm} on telecom-specific datasets and to enhance their performance in network security tasks. This is particularly important for future `\gls{ai}-native’ networks, where \gls{ai} needs to detect and identify threats autonomously and with the highest accuracy. 

%  In this work, we present the first study (to the best of our knowledge) to explore and investigate the suitability of \glspl{llm} for network security evaluation.

% \color{red}
% \begin{enumerate}
%     % \item \textcolor{red}{Providing examples to the \glspl{llm} improves their performance. This is inline with~\cite{brown2020language}.} \textcolor{red}{advantage of FS:  The advantage of such an approach is that it provides the \gls{llm} with task-specific examples which help the model's performance~\cite{brown2020language}}

%     % \item \textcolor{red}{training of LLMs on telecom-specific datasets to enhance their performance,}

%     % \item \textcolor{red}{LLMs may not always consider the threat classifications as a result of the second-order effect. This will limit the threat identification and may eventually result in not identifying all the possible risks associated with a threat.}
%     \item \textcolor{red}{In future `AI-native' and `zero-touch' networks, AI needs to detect and identify threats with higher accuracy.}
%     \item in general, the LLMs showed bad performance even with the internet search, which indicates that the LLMs were unable to extract the stride classification information from the published literature and the standards.
%     \item The maximum accuracy achieved is $72\%$, so there is scope for performance enhancements.
% \end{enumerate}

\section*{Acknowledgment}
This work was supported in part by the Natural Sciences and Engineering Research Council of Canada (NSERC). 

\bibliographystyle{IEEEtran} 
\bibliography{library}

\end{document}

%% file: acronym.tex
\newacronym{1g}{1G}{First Generation}
\newacronym{2g}{2G}{Second Generation}
\newacronym{3g}{3G}{Third Generation}
\newacronym{4g}{4G}{Fourth Generation}
\newacronym{5g}{5G}{Fifth Generation}
\newacronym{5g-advanced}{5G-Advanced}{5G-Advanced}
\newacronym{6g}{6G}{Sixth Generation}

\newacronym{vnf}{VNF}{Virtualized Network Function}

\newacronym{mmtc}{mMTC}{massive Machine Type Communications}
\newacronym{embb}{eMBB}{enhanced Mobile Broadband}
\newacronym{urllc}{URLLC}{Ultra-Reliable Low Latency Communications}
\newacronym{sdn}{SDN}{Software Defined Networking}
\newacronym{nfv}{NFV}{Network Function Virtualization}
\newacronym{nf}{NF}{Network Function}
\newacronym{dn}{DN}{Data Network}
\newacronym{arpf}{ARPF}{Authentication credential Repository and Processing Function}

\newacronym{pfcp}{PFCP}{Packet Forwarding Control Protocol}
\newacronym{udp}{UDP}{User Datagram Protocol}

\newacronym{supi}{SUPI}{Subscription Permanent Identifier}
\newacronym{capex}{CAPEX}{Capital Expenditure}
\newacronym{opex}{OPEX}{Operational Expenditure}
\newacronym{dos}{DoS}{Denial-of-Service}
\newacronym{ddos}{DDoS}{Distributed Denial-of-Service}
\newacronym{minlp}{MINLP}{Mixed-Integer Nonlinear Programming}
\newacronym{ilp}{ILP}{Integer Linear Programming}
\newacronym{upf}{UPF}{User Plane Function}
\newacronym{mec}{MEC}{Multi-access Edge Computing}
\newacronym{qos}{QoS}{Quality of Service}
\newacronym{sfc}{SFC}{Service Function Chain}
\newacronym{pdu}{PDU}{Protocol Data Unit}
\newacronym{ns}{NS}{Network Slicing}
\newacronym{sla}{SLA}{Service Level Agreement}
\newacronym{dps}{DPS}{Data Plane Services}
\newacronym{cps}{CPS}{Control Plane Services}
\newacronym{vm}{VM}{Virtual Machine}
\newacronym{mip}{MIP}{Mixed Integer Programming}

\newacronym{sba}{SBA}{Service-Based Architecture}

\newacronym{amf}{AMF}{Access and Mobility Function}
\newacronym{smf}{SMF}{Session Management Function}
\newacronym{nrf}{NRF}{Network Repository Function}
\newacronym{scip}{SCIP}{Solving Constraint Integer Programs}
\newacronym{ue}{UE}{User Equipment}

\newacronym{hplmn}{H-PLMN}{Home \gls{plmn}}
\newacronym{plmn}{PLMN}{Public Land Mobile Network}
\newacronym{eap-aka}{EAP-AKA'}{Extensible Authentication Protocol-Authentication and Key Agreement}
\newacronym{aka}{AKA}{Authentication and Key Agreement}
\newacronym{gnb}{gNB}{gNodeB}
\newacronym{3gpp}{3GPP}{3rd Generation Partnership Project}
\newacronym{ran}{RAN}{Radio Access Network}
\newacronym{udm}{UDM}{Unified Data Management}
\newacronym{ausf}{AUSF}{Authentication Server Function}

\newacronym{kpi}{KPI}{Key Performance Indicator}

\newacronym{llm}{LLM}{Large Language Model}
\newacronym{genai}{GenAI}{Generative Artificial Intelligence}
\newacronym{ai}{AI}{Artificial Intelligence}
\newacronym{guti}{GUTI}{Globally Unique Temporary Identity}

\newacronym{imei}{IMEI}{International Mobile Equipment Identity}
\newacronym{s-nssai}{S-NSSAI}{Single Network Slice Selection Assistance Information} 
\newacronym{mitm}{MiTM}{Man-in-The-Middle}

\newacronym{stride}{STRIDE}{Spoofing, Tampering, Repudiation, Information Disclosure, Denial of Service, and Elevation of Privilege}

\newacronym{ltm}{LTM}{Large Telecom Model}

\newacronym{nssaa}{NSSAA}{Network Slice-Specific Authentication and Authorization}

\newacronym{zs}{ZS}{Zero-Shot}
\newacronym{fs}{FS}{Few-Shot}